\documentclass[pre,a4paper,showpacs,preprint,byrevtex]{revtex4}
\usepackage{graphicx}
\usepackage{dcolumn}
\usepackage{amsmath}
\usepackage{bm}
\usepackage{array}

\begin{document}

\title{Quark self-energy and relativistic flux tube model}

\author{Fabien \surname{Buisseret}}
\thanks{FNRS Research Fellow}
\email[E-mail: ]{fabien.buisseret@umh.ac.be}
\author{Claude \surname{Semay}}
\thanks{FNRS Research Associate}
\email[E-mail: ]{claude.semay@umh.ac.be}
\affiliation{Groupe de Physique Nucl\'{e}aire Th\'{e}orique,
Universit\'{e} de Mons-Hainaut,
Acad\'{e}mie universitaire Wallonie-Bruxelles,
Place du Parc 20, B-7000 Mons, Belgium}

\date{\today}

\begin{abstract}
The contribution of the quark self energy to the meson masses is studied
in the framework of the relativistic flux tube model. The equivalence
between this phenomenological model and the more QCD based rotating
string Hamiltonian is used as a guide to perform the calculations. It is
shown that the addition of the quark self energy to the relativistic
flux tube model preserves the linearity of the Regge trajectories. But,
following the definition taken for the constituent quark masses, the
Regge slope is preserved or decreased. In this last case, experimental
data can only by reproduced by using a string tension around
$0.245$~GeV$^2$. Two procedures are also studied to treat the pure flux
tube contribution as a perturbation of a spinless Salpeter Hamiltonian.
\end{abstract}


\maketitle

\section{Introduction}
\label{sec:intro}

A successful way of understanding the properties of the mesons is to
approximate the gluon exchanges between the quark and the antiquark by a
string (the QCD string), characterized by its energy density, or
tension. The relativistic flux tube model (RFTM) is a phenomenological
model based on this picture \cite{laco89, tf_2}. More recently, the
rotating string model (RSM) has been derived from the Nambu-Goto
Lagrangian as an effective model which also describes a meson as a quark
and an antiquark linked by a string \cite{dubi94,morg99}. The physical
content of these models is very similar, and it has been shown that they
are actually classically equivalent once the auxiliary fields appearing
in the RSM are completely eliminated \cite{sema04,buis042}. A rather
good description of the experimental meson spectrum can be obtained with
the original RFTM supplemented by a Coulomb term \cite{sema95}. But
unfortunately, a strong coupling constant larger than what it is
expected
from experimental analysis and lattice simulations must be considered.
The necessity of finding new contributions arising from neglected
physical mechanisms is thus clear. Recently, a quark self energy (QSE)
contribution was introduced, which is due to the color magnetic moment
of the quark propagating through the vacuum \cite{Sim1}. The QSE brings
a negative contribution to the hadron masses, and seems to be an
interesting way of reproducing the experimental data with a smaller
Coulomb term.

Our purpose here is to study the influence of the QSE on the meson
masses, especially using the RFTM. Our paper is organized as follows.
The Sec.~\ref{sec:rs_rft} is a short presentation of both classical RFTM
and RSM, where we also underline their classical equivalence. Since the
equations describing these models are quite complicated, approximate
equations are developed in Sec.~\ref{sec:str_pert}, from which an
analytical mass formula is derived in Sec.~\ref{sec:massform}. In
Sec.~\ref{sec:compar}, the quality of the approximate equations are
compared with the original quantized RFTM. We then introduce the QSE in
Sec.~\ref{sec:qse} and discuss its effects on the meson spectrum in
Sec.~\ref{sec:qse_add}. A comparison with experimental data is performed
in Sec.~\ref{sec:exp_dat}. Finally, some concluding remarks are outlined
in Sec.~\ref{sec:conclu}.

\section{Rotating string and relativistic flux tube}
\label{sec:rs_rft}

We will present in this section two effective meson models : the
rotating string model (RSM) and the relativistic flux tube model (RFTM).
Our purpose is to give a presentation of the principal features of these
models, and to underline their classical equivalence.

It has been shown in Ref.~\cite{dubi94} that starting from the QCD
Lagrangian, the Lagrange function of a meson with spinless quarks can be
built from the Nambu-Goto
action. For two quarks with masses $m_{1}$ and $m_{2}$, and a string
with tension $a$, this action has the well-known form
\begin{eqnarray}
\label{nambu}
{\cal
L}=-m_{1}\sqrt{\dot{\bm x}^{2}_{1}}-m_{2}\sqrt{\dot{\bm
x}^{2}_{2}}
-a\int^{1}_{0}d\beta\sqrt{(\dot{\bm w}\bm w')^{2}-\dot{\bm w}^{2}{\bm
w'}^{2}}.
\end{eqnarray}
In this action, $\bm x_i$ is the coordinate of quark $i$ and $\bm w$ is
the coordinate of the string. $\bm w$ depends
on two variables defined on the string worldsheet : one is spacelike
$\beta$
and the other timelike $\tau$. We have also defined
$\bm w'=\partial_{\beta} \bm w$ and
$\dot{\bm w}=\partial_{\tau} \bm w$. Introducing auxiliary fields
(also known as einbein fields) to get rid of the square root in
(\ref{nambu})  and making the straight line ansatz to describe the
string connecting the quark and the antiquark, an effective Hamiltonian
can be derived, which reads \cite{guba94}
\begin{eqnarray}
\label{QCD_eq1}
H(\mu_{i},\nu)&=&\frac{1}{2}\left[\frac{p^2_r+m^{2}_{1}}{\mu_{1}}+
\frac{p^{2}_{r}
+m^{2}_{2}}{\mu_{2}}+\mu_{1}+\mu_{2} \right.\nonumber \\
&&\left.+ a^{2}r^{2}\int^{1}_{0}\frac{d\beta}{
\nu}+ \int^{1}_{0}d\beta\nu+\frac{L^{2}/r^{2}}{[\mu_{1}(1-\zeta)^{2}+
\mu_{2}\zeta^{2}+\int^{1}_{0} d\beta\,
(\beta-\zeta)^{2}\, \nu] } \right].
\end{eqnarray}
$p_r$ is the common radial quark momentum.
The parameter $\zeta$ defines the position $\bm R$ of the center of
mass: $\bm R=\zeta \bm x_1+(1-\zeta) \bm x_2$, and $L$ is
the orbital angular momentum of the system. The auxiliary
fields $\mu_{1}$ and $\mu_{2}$ are seen as effective masses of the
quarks whose current masses are $m_{1}$ and $m_{2}$. The last
auxiliary field, $\nu$, can be interpreted  in the same way as an
effective energy for the string whose ``static" energy
is $a r$. One can get rid of these auxiliary fields by a variation of
the Hamiltonian~(\ref{QCD_eq1}). Their extremal values, denoted as
$\mu_{i\, 0}$ and $\nu_{0}$, are the solutions of
\begin{subequations}
\label{muelim}
\begin{equation}
\label{muelima}
\left.\frac{\delta H(\mu_{i},\nu)}{\delta \mu_{i}}\right|_{\mu_{i}=\mu_{
i\, 0}} =0,
\end{equation}
\begin{equation}
\label{muelimb}
\left.\frac{\delta H(\mu_{i},\nu)}{\delta \nu}\right|_{\nu=\nu_{0}}=0.
\end{equation}
\end{subequations}

The equations of the RSM can be derived from the
Hamiltonian~(\ref{QCD_eq1}) by the elimination of $\nu$ thanks to the
condition~(\ref{muelimb}) \cite{morg99}. We
will consider in this paper the symmetrical case, where $m_{1}=m_{2}=m$
and $\mu_{1}=\mu_{2}=\mu$:
\begin{subequations}
\label{rs23}
\begin{equation}\label{rs2}
\frac{\sqrt{\ell(\ell+1)}}{ar^{2}}=\frac{\mu y}{ar}+\frac{1}{4y^{2}}(
\arcsin y-y\sqrt{1-y^{2}}),
\end{equation}
\begin{equation}\label{rs3}
H^{\text{RS}}(\mu)=\frac{p^{\, 2}_{r}+m^{2}}{\mu}+\mu+\frac{ar}{y}
\arcsin
y+\mu y^{2}.
\end{equation}
\end{subequations}
In the general case, where $m_{1}\neq m_{2}$, a third equation has to be
taken into account, which expresses the cancellation of the total
momentum
in the center of mass frame \cite{buis042}. If one considers $\mu$ as a
number, one can directly solve Eqs.~(\ref{rs23}) and
find the meson mass, after a minimization of this mass with respect to
$\mu$ \cite{morg99}.

\par Since the RSM equations contain the remaining auxiliary field
$\mu$ and a variable $y$ whose physical interpretation is not a priori
clear, it appears interesting to go a step further and to get rid of
$\mu$. As it is shown in Ref.~\cite{sema04}, the elimination of $\mu$
with the condition (\ref{muelim}) leads to the extremal value
\begin{equation}\label{muextr}
\mu_{0}=\sqrt{\frac{p^{2}_{r}+m^{2}}{1-y^{2}}}.
\end{equation}
Moreover, the replacement of $\mu$ by $\mu_{0}$ in Eqs.~(\ref{rs23})
leads to the following expressions
\begin{subequations}
\label{tf12}
\begin{equation}\label{tf1}
\frac{L}{ar^{2}}=\frac{1}{ar}v_{\bot}\gamma_{\bot} W_{r}+f(v_{\bot}),
\end{equation}
\begin{equation}\label{tf2}
H^{\text{RS}}(\mu_{0})=H^{\text{RFT}}=2\gamma_{\bot} W_{r}+ar\frac{
\arcsin
v_{\bot}}{v_{\bot}},
\end{equation}
\end{subequations}
where we have defined
\begin{equation}
\label{gammabot}
y=v_{\bot},
\quad \gamma_{\bot}=\frac{1}{\sqrt{1-v^{2}_{\bot}}},
\quad f(v_{\bot})=\frac{\arcsin v_{\bot} }{4v^{2}_{\bot}}-\frac{1}{4v_{
\bot} \gamma_{\bot}}
\quad \text{and} \quad W_{r}=\sqrt{p^{2}_{r}+m^{2}}.
\end{equation}
Eqs.~(\ref{tf12}) are precisely those of the RFTM as
they appear in Ref.~\cite{laco89}. The mysterious variable $y$ is now
simply interpreted as the transverse velocity $v_{\bot}$ of the quarks,
and the physical content of $\mu$ is clarified. Using
definitions~(\ref{gammabot}), we can rewrite (\ref{muextr}) in the form
\begin{equation}\label{murft}
\mu_0=W_{r}\gamma_{\bot}.
\end{equation}
\par Originally, the RFTM was built on phenomenological arguments. But
our derivation of the RFTM shows that it can be derived from the Nambu-
Goto Lagrangian. We also clearly see the equivalence of the RSM and
RFTM, when the auxiliary fields are eliminated. Let us note that this
equivalence is also true in general, with $m_{1}\neq m_{2}$
\cite{buis042}.

By application of the usual correspondence rules
\begin{equation}
p^{2}_{r}\rightarrow -\frac{1}{r}\frac{\partial^{2}}{\partial r^{2}}r
\quad \text{and} \quad L\rightarrow\sqrt{\ell(\ell+1)},
\end{equation}
the quantized equations of the RFTM are given by \cite{laco89}
\begin{subequations}
\label{tf_equa}
\begin{equation}\label{tf_equa1}
\frac{2\sqrt{\ell(\ell+1)}}{r}=\{v_{\bot}\gamma_{\bot},W_{r}\}+a\{r,f(v_
{\bot})\},
\end{equation}
\begin{equation}\label{tf_equa2}
H^{\text{RFT}}=\{\gamma_{\bot},W_{r}\}+\frac{a}{2}\{r,\frac{\arcsin v_{
\bot}}{v_{\bot}}\}.
\end{equation}
\end{subequations}
The anticommutators $\{A, B\}=AB+BA$ arise because $v_{\bot}$, $r$, and
$p_{r}$ are non commuting operators.

The quantized equations of the RSM, which are not used here, are given
in Ref.~\cite{buis041}. The diagonalization of Hamiltonian
$H^{\text{RFT}}$
directly provides the physical masses $M^{\text{RFT}}$.
Equations~(\ref{tf_equa}) can be numerically solved, as it is done
in Refs.~\cite{sema95,buis041}. Supplemented by an appropriate short
range potentials, like a Coulomb term, the quantized RFTM can rather
well reproduce the meson spectra \cite{sema95}.

\section{String as a perturbation}
\label{sec:str_pert}

When $\ell=0$, the RFTM reduces to a spinless Salpeter Hamiltonian
(SSH) with a linear confinement potential $ar$ \cite{laco89}. If $\ell$
is small, the contribution of the string is also small, and then it can
be treated as a perturbation of the SSH. We will show that the
contribution of the string can be obtained by two different procedures,
that lead to different definition of the auxiliary field $\mu_0$.

Let us start with the RSM and
consider that the transverse velocity of the quarks is small: $y\ll 1$.
We can develop formulas~(\ref{rs23}) at the second order
in $y$. We obtain then
\begin{subequations}
\begin{equation}\label{y2def}
y^{2}\approx\frac{\ell(\ell+1)}{r^{2}\left(\frac{ar}{6}+\mu\right)^{2}},
\end{equation}
\begin{equation}\label{RSap}
H^{\text{RS}}(\mu)\approx\frac{p^{2}_{r}+m^{2}}{\mu}+\mu+ar+\left(\frac{
ar}{6}+\mu  \right)y^{2}.
\end{equation}
\end{subequations}
Inserting (\ref{y2def}) in (\ref{RSap}), and introducing
$\vec p\,^2=p^2_r+\ell(\ell+1)/r^2$, we can write down an
approximate Hamiltonian
\begin{equation}\label{ham_a}
H^{\text{A}}(\mu)=\frac{\vec p\,^2+m^{2}}{\mu}+\mu+ar-\frac{a\ell(\ell
+1)}{r\mu(6\mu+ar)},
\end{equation}
defining what we call here the perturbative flux tube model (PFTM).
We see that $H^{\text{A}}(\mu)$ is the sum of an usual SSH with linear
confinement in the auxiliary field formalism \cite{sema04}
\begin{equation}
\label{hss}
H^{\text{SS}}(\mu)=\frac{\vec p\,^2+m^{2}}{\mu}+\mu+ar,
\end{equation}
and a specific contribution of the string, which reads
\begin{equation}\label{strcorr_debut}
\Delta H_{\text{str}}(\mu)=-\frac{a\ell(\ell+1)}{r\mu(6\mu+ar)}.
\end{equation}
Since we want to treat $\Delta H_{\text{str}}(\mu)$ as a perturbation,
we
will eliminate $\mu$ with the condition~(\ref{muelima}) applied for the
Hamiltonian $H^{\text{SS}}(\mu)$. This leads to the extremal value
\begin{equation}
\label{muss}
\mu_0=\sqrt{\vec p\,^2+m^{2}},
\end{equation}
and after replacement in (\ref{hss}), to the spinless Salpeter
Hamiltonian
\begin{equation}
\label{hss2}
H^{\text{SS}}=2\sqrt{\vec p\,^2+m^{2}}+ar.
\end{equation}
Let us note that formula~(\ref{muss}) is different from the extremal
value~(\ref{murft}).
The string correction to compute is then given by \cite{bada02}
\begin{equation}
\label{strcorr1}
\Delta M_{\text{str}}=\langle \Delta H_{\text{str}}(\mu_0)\rangle
-\frac{a\ell(\ell+1)\langle 1/r\rangle}{\langle
\sqrt{\vec p\,^2+m^{2}} \rangle\left(6\langle \sqrt{\vec p\,^2+m
^{2}} \rangle+a\langle r\rangle\right)},
\end{equation}
in which the mean value is performed with an eigenstate of
$H^{\text{SS}}$. This string contribution was previously obtained in
Ref.~\cite{bada02}, where it is shown that its accuracy is better than
3\%.

We can now
invert the order of the operations: Firstly to eliminate $\mu$ in the
RSM and obtain the RFTM, then make the same approximation as before.
When $v_{\bot}\ll 1$, the RFTM equations
can be developed at the second order in $v_{\bot}$ and we have
\begin{subequations}
\begin{equation}\label{tfapp1}
v_{\bot}\approx\frac{\sqrt{\ell(\ell+1)}}{r\left(W_{r}+\frac{ar}{6}
\right)},
\end{equation}
\begin{equation}\label{tfapp2}
H^{\text{RFT}}\approx2W_{r}+ar+
v^{2}_{\bot}\left(W_{r}+\frac{ar}{6}\right).
\end{equation}
\end{subequations}
Replacing Eq.~(\ref{tfapp1}) in Eq.~(\ref{tfapp2}) leads to the
Hamiltonian
\begin{equation}\label{htildea}
\tilde{H}^{\text{A}}=2W_{r}+ar+\frac{\ell(\ell+1)}{r^{2}\left(W_{r}+
\frac{
ar}{6}\right)}.
\end{equation}
Using the fact that
\begin{equation}
\sqrt{\vec p\,^2+m^{2}}=\sqrt{W^{2}_{r}+\frac{\ell(\ell+1)}{r^{2}}},
\end{equation}
we can rewrite Hamiltonian~(\ref{htildea}) in the form
\begin{equation}
\tilde{H}^{\text{A}}=2\sqrt{\vec p\,^2+m^{2}}+ar+2W_{r}-2\sqrt{W^{2}_{
r}+\frac{\ell(\ell+1)}{r^{2}}}+\frac{\ell(\ell+1)}{r^{2}\left(W_{r}+
\frac{ar}{6}\right)}.
\end{equation}
Now, we make a new approximation and assume that
$W_{r}\gg \sqrt{\ell(\ell+1)}/r$. In this case, which is justified in
the limit $v_{\bot}\ll 1$, a first order expansion leads to
\begin{equation}
\sqrt{W^{2}_{r}+\frac{\ell(\ell+1)}{r^{2}}}\approx W_{r}+\frac{\ell(\ell
+1)}{2r^{2}W_{r}}.
\end{equation}
We finally get from $\tilde{H}^{\text{A}}$ the Hamiltonian
\begin{equation}\label{happtffin}
H^{\text{A}}= 2\sqrt{\vec p\,^2+m^{2}}+ar-\frac{a\ell(\ell+1)}{rW_{r}(
6W_{r}+ar)}.
\end{equation}
The approximate Hamiltonian~(\ref{happtffin}) can again be seen as
the sum of the Hamiltonian $H^{\text{SS}}$ (\ref{hss2})
and a contribution of the flux tube
\begin{equation}
\Delta H_{\text{rft}}=-\frac{a\ell(\ell+1)}{rW_{r}(6W_{r}+ar)}.
\end{equation}
As for formula~(\ref{strcorr1}), the flux tube contribution is given by
\begin{equation}\label{strcorr2}
\Delta M_{\text{rft}}=\langle \Delta H_{\text{rft}}\rangle=-\frac{a\ell(
\ell
+1)\langle 1/r\rangle }{\langle \sqrt{p^{2}_{r}+m^{2}} \rangle\left(6
\langle \sqrt{p^{2}_{r}+m^{2}} \rangle+a\langle r\rangle\right)},
\end{equation}
in which the mean value is again performed with an eigenstate of
$H^{\text{SS}}$. We immediately see that formulas~(\ref{strcorr1}) and
(\ref{strcorr2}) are
different. In Sec.~\ref{sec:compar}, we will study the qualities of both
corrections.

\section{A mass formula}
\label{sec:massform}

In the following, we will specially focus on the massless case, $m=0$,
for which we can expect the largest contributions of the flux tube
($\mu$ increases with the quark mass). In this case, it is possible to
find an approximate analytical mass formula. This will allow a better
understanding of the effects of the QSE
(see Sec.~\ref{sec:Reg_t}). Starting from the Hamiltonian~(\ref{ham_a}),
we define dimensionless conjugate variables $\vec{x}$ and $\vec{q}$
by the following scaling
\begin{equation}
\vec{r}=\frac{\vec{x} }{(\mu a)^{1/3}} \quad \text{and} \quad
\vec{p}=(\mu a)^{1/3}\vec{q}.
\end{equation}
In this section, we treat the auxiliary field $\mu$ as a number
\cite{Sim1,bada02}.
We can write the Hamiltonian~(\ref{ham_a}) in the form
\begin{equation}
\label{HSSnull2}
H^{\text{SS}}(\mu)=\left(\frac{a^2}{\mu}\right)^{1/3}\left(\vec q\,^2+
x\right)+\mu,
\end{equation}
with $x=\left|\vec x\,\right|$. Its eigenvalues are
consequently
\begin{equation}\label{e_ka}
M^{\text{SS}}(\mu; n\ell)=\left(\frac{a^2}{\mu}\right)^{1/3}
\epsilon_{n\ell}+\mu,
\end{equation}
where $\epsilon_{n\ell}$ is an eigenvalue of the Hamiltonian
$\left(\vec q\,^2+x\right)$, which is easy to solve numerically.
Simple analytical approximate expressions of $\epsilon_{n\ell}$ can be
found
for $\epsilon_{n0}$ \cite{sema04} and $\epsilon_{0\ell}$ \cite{luch95}
\begin{subequations}
\begin{equation}
\epsilon_{n0}\approx\left[\frac{3\pi}{4}\left(2n+\frac{3}{2}\right)
\right]^{2/3},
\end{equation}
\begin{equation}
\epsilon_{0\ell}\approx\frac{3}{2^{2/3}}\left(\ell+\frac{3}{2}\right)^{1
/3}\left[\frac{\Gamma(\ell+2)}{\Gamma\left(\ell+\frac{3}{2}\right)}
\right]^{2/3}.
\end{equation}
\end{subequations}
A more complicated approximate formula exists also in the general case
\cite{bose76}.

Assuming that the quantities $\epsilon_{n\ell}$ are known, we can
compute the extremal value of $\mu$ by a minimization of
relation~(\ref{e_ka}), which leads to
\begin{equation}\label{mu0}
\mu_{0\, n\ell}=\sqrt{a} \left(\frac{\epsilon_{n\ell}}{3}\right)^{3/4},
\end{equation}
and
\begin{equation}
M^{\text{SS}}_{n\ell}(\mu_{0\, n\ell})=4\mu_{0\, n\ell}.
\end{equation}
We will now drop the explicit dependence in $n$ and $\ell$ of the
different terms to simplify the notations. The perturbation
theory tells us that the total energy is
\begin{equation}\label{e_A}
M^{\text{A}}(\mu_{0})=4\mu_{0}-\frac{a\ell(\ell+1)}{\langle
r\rangle\mu_{0}(6\mu_{0}+a\langle r\rangle)},
\end{equation}
where $\langle 1/r \rangle$ is replaced by $1/\langle r\rangle$.
The next step is to use the Hellmann-Feynman theorem
\cite{sema04,feyn1},
which states that
\begin{equation}\label{meanv}
M^{\text{SS}}(\mu_{0})=4\mu_{0}=2\mu_{0}+a\langle r\rangle.
\end{equation}
Extracting $\langle r\rangle$ from relation~(\ref{meanv}) and replacing
it in relation~(\ref{e_A}), we finally obtain the mass formula
\begin{equation}\label{massform}
M^{\text{A}}(\mu_{0})\approx4\mu_{0}-\frac{a^{2}\ell(\ell+1)}{16\mu^{3}_
{0
}},
\end{equation}
where $\mu_{0}$ is given by formula~(\ref{mu0}).

\par The mass formula (\ref{massform}), even approximate, exhibits
Regge trajectories. This can be easily checked when $n=0$. Using the
fact that
\begin{equation}\label{eps_bigl}
\lim_{\ell \rightarrow \infty }\epsilon_{0\ell}\approx 3\left(\frac{\ell
+\frac{3}{2}}{2}\right)^{2/3},
\end{equation}
with formula~(\ref{mu0}), we obtain the following expression
for the extremal value of the auxiliary field at large values of $\ell$
\begin{equation}\label{mumassform}
\mu_{0}\approx\sqrt{\frac{a\ell}{2}}.
\end{equation}
Replacing Eq.~(\ref{mumassform}) into Eq.~(\ref{massform}), we get
\begin{equation}
\left(M^{\text{A}}\left(\mu_{0}\right)\right)^{2}\approx \frac{225}{32}a
\ell.
\end{equation}
This linear relation between the squared mass and the angular momentum
reproduces qualitatively the Regge trajectories. The Regge slope is here
$7.03\, a$, a higher value than the one predicted by the RFTM, which
gives a slope equal
to $2\pi a$ \cite{laco89}. Finally, we can observe that
\begin{equation}
\lim_{\ell \rightarrow \infty }
\frac{\frac{a^{2}\ell(\ell+1)}{16\mu^{3}_{0}}}{4\mu_{0}} = \frac{1}{16}.
\end{equation}
When $\ell=0$, this ratio is vanishing. This justifies to treat the
contribution of the string as a perturbation.

\section{Comparison between exact and approximate string contribution}
\label{sec:compar}

Before studying the contribution of the QSE to the RFTM, it is
interesting to examine the relevance of the approximate treatment for
the
flux tube developed in Sec.~\ref{sec:str_pert}. For this purpose, we
compare here some masses computed with the ``exact" RFTM by numerically
solving Eqs.~(\ref{tf_equa}), as it is done for example in
Ref.~\cite{buis041}, with masses computed in the framework of the PFTM
with both string corrections~(\ref{strcorr1}) and (\ref{strcorr2}).

The symbol $M_{E}$ designs a mass computed with the RFTM, and
$M_{P1}$, $M_{P2}$ the corresponding PFTM masses evaluated with the
string corrections (\ref{strcorr1}) and (\ref{strcorr2})
respectively. The quantities
\begin{equation}
\Delta M_{E,Pi}=\left|\frac{M_{E}-M_{Pi}}{M_{E}} \right|
\end{equation}
measure the differences between the RFTM and its approximations.

\begin{table}[h]
\caption{$\Delta M_{E,P1}$ (correction~(\ref{strcorr1})) in $\%$, for
different states, with $m=0$.
$\ell$ is the orbital angular momentum of the state and $n$ is the
number of nodes at finite distance.}
\label{tab:DeltaM1}
\begin{ruledtabular}
\begin{tabular}{rccccc}
$\ell$ & $1$ & $2$ & $3$ & $4$ & $5$ \\
\hline
$n=0$ & $3.26$ & $3.78$ & $4.01$ & $4.17$ & $4.29$\\
$1$ & $1.08$ & $1.31$ & $1.53$ & $1.74$ & $1.94$\\
$2$ & $0.68$ & $0.86$ & $1.04$ & $1.22$ & $1.40$\\
$3$ & $0.46$ & $0.56$ & $0.68$ & $0.81$ & $0.95$\\
$4$ & $0.36$ & $0.43$ & $0.52$ & $0.64$ & $0.76$\\
$5$ & $0.29$ & $0.35$ & $0.40$ & $0.49$ & $0.58$\\
\end{tabular}
\end{ruledtabular}
\end{table}

Table~\ref{tab:DeltaM1} presents the difference between the RFTM
and the PFTM with correction~(\ref{strcorr1}). As the basis of the
approximation was to consider a small $v_{\bot}$, it is not surprising
to
see that $\Delta M_{E,P1}$ increases with $\ell$. On the contrary,
$\Delta M_{E,P1}$ decreases for an increasing radial quantum number $n$.
This can be understood  by the presence of the operator $p^{2}_{r}$ in
the
denominator of Eq.~(\ref{strcorr1}). This term becomes larger with $n$,
and makes the contribution of the string smaller, leading to a
decreasing
of $\Delta M_{E,P1}$. Globally, the approximation considered in
Table~\ref{tab:DeltaM1} is rather good, in particular when $n\geq\ell$.

\begin{table}[h]
\caption{Same as in Table~\ref{tab:DeltaM1} but for $\Delta M_{E,P2}$
(correction~(\ref{strcorr2})).}
\label{tab:DeltaM2}
\begin{ruledtabular}
\begin{tabular}{rccccc}
$\ell$ & $1$ & $2$ & $3$ & $4$ & $5$\\
\hline
$n=0$ & $1.03$ & $7.81$ & $15.21$ & $22.52$ & $29.54$\\
$1$ & $0.44$ & $0.88$ & $2.55$ & $4.41$ & $6.36$\\
$2$ & $0.40$ & $0.67$ & $0.73$ & $1.62$ & $2.64$\\
$3$ & $0.35$ & $0.14$ & $0.23$ & $0.72$ & $1.30$\\
$4$ & $0.31$ & $0.19$ & $0.02$ & $0.30$ & $0.66$\\
$5$ & $0.26$ & $0.19$ & $0.06$ & $0.13$ & $0.37$\\
\end{tabular}
\end{ruledtabular}
\end{table}

Table~\ref{tab:DeltaM2} presents the difference between the RFTM and the
PFTM with correction~(\ref{strcorr2}). When $n=0$, we immediately see
that this approximation is not so good, because $\Delta M_{E,P2}$
becomes large very quickly with $\ell$. The situation is better for
larger
values of $n$, when our approximation $W_{r}\gg \sqrt{\ell(\ell+1)}/r$,
is particularly justified. But the evolution is less monotonic than in
Table~\ref{tab:DeltaM1}. In particular, for a fixed value of $n$,
$\Delta M_{E,P2}$ decreases to a minimal value, and then increases again
with $\ell$. In conclusion, in the framework of the PFTM, the
correction~(\ref{strcorr1}), proposed in Ref.~\cite{bada02}, seems
preferable (even if the other one can sometimes give a better result).
It provides quite globally good results.

\section{The quark self-energy}
\label{sec:qse}

\subsection{Definition}\label{sec:qsedef}
Recently, it was shown that the QSE contribution, which is created by
the color magnetic moment of the quark propagating through the vacuum
background field, adds a negative constant to the hadron masses \cite{
Sim1}. Its negative
sign is due to the paramagnetic nature of the particular mechanism at
work in this case. Using the Fock-Feynman-Schwinger representation of
the quark
Green's function, one can obtain the QSE contribution as a shift of the
squared mass of the quark \cite{Sim1,DiGia1} which reads
\begin{equation}\label{Dmdef}
\Delta m^{2}=-3m\int^{\infty}_{0}dz\, z^{2} K_{1}(mz) \left(D(z) +
D_{1}(z)\right),
\end{equation}
where $D$ and $D_{1}$ are quark correlators and $K_{1}$ the Mac-Donald
function. The properties of these correlators were studied by lattice
simulations in the quenched case \cite{DiGia2}. One has then
\begin{equation}\label{Ddef}
D(z)\approx 3 D_{1}(z)= D(0) \exp(-\left|z\right|\delta),
\end{equation}
with $\delta=1/T_{g}$. $T_{g}$ is the gluonic correlation length, whose
value is estimated at about 0.15-0.2~fm. This locates $\delta$ in the
interval 1.0-1.3~GeV. The results (\ref{Ddef}) allow us to find an
analytic form for the integral (\ref{Dmdef})
\begin{equation}
\Delta m^{2}=-4mD(0)\varphi(m/\delta),
\end{equation}
where, defining $\epsilon=m/\delta$, we can write $\varphi(\epsilon)$ as
\begin{equation}
\varphi(\epsilon)=\left\{
\begin{array}{lll}
&\dfrac{1}{\delta^{3}}\left[\dfrac{-3\epsilon}{\left(1-\epsilon^{2}
\right)
^{5/2}}\ln\left(\dfrac{1+\sqrt{1-\epsilon^{2}}}{\epsilon}\right)+\dfrac{
1+
2\epsilon^{2}}{\epsilon\left(1-\epsilon^{2})^{2}\right)}\right]=\dfrac{1
}{\delta^{3}}\phi_{1}(\epsilon) &(\epsilon < 1)\\
&
\dfrac{1}{\delta^{3}}\left[\dfrac{-3\epsilon}{\left(\epsilon^{2}-1\right
)^
{5/2}}\arctan\left(\sqrt{\epsilon^{2}-1}\right)+\dfrac{1+2\epsilon^{2}}{
\epsilon\left(1-\epsilon^{2})^{2}\right)}\right]=\dfrac{1}{\delta^{3}}
\phi_{2}(\epsilon) &(\epsilon > 1) .
\end{array} \right.
\end{equation}
One can check that $\phi_{1}(1)=\phi_{2}(1)=2/5$, and that
\begin{equation}
\lim_{\epsilon\rightarrow0}\phi_{1}(\epsilon)=\frac{1}{\epsilon}.
\end{equation}

\begin{figure}[h]
\begin{center}
\includegraphics*[width=8.0cm]{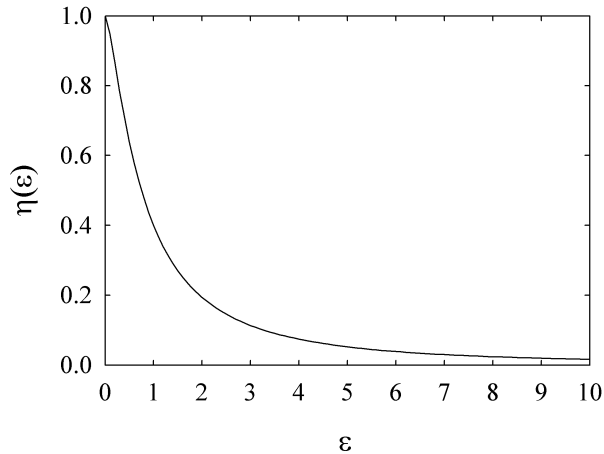}
\end{center}
\caption{Plot of $\eta(\epsilon)$.}
\label{fig:eta}
\end{figure}

For a purely exponential correlator, as it is the case here, $D(0)$ is
connected with the string tension $a$ by the relation
\begin{equation}
a=\frac{1}{2}\int d^{2}x\, D(x)=\frac{\pi D(0)}{\delta^{2}},
\end{equation}
so we find
\begin{equation}\label{Dm2}
\Delta m^{2}=-\frac{4am\delta^{2}}{\pi}\varphi(\epsilon).
\end{equation}
For convenience, we define a new dimensionless function,
$\eta(\epsilon)$, by
\begin{equation}\label{etadef}
\eta(\epsilon)=\delta^{3}\epsilon\varphi(\epsilon).
\end{equation}
We see in Fig.~\ref{fig:eta} that $\eta(0)=1$
and that $\eta(\epsilon)$ rapidly decreases for increasing values of
$\epsilon$.

Thanks to the definition (\ref{etadef}), Eq.~(\ref{Dm2}) takes its
final form \cite{Sim1}
\begin{equation}\label{qsedef}
\Delta m^{2}=-\frac{4a}{\pi}\eta(\epsilon).
\end{equation}
An important
ingredient we have used to get the contribution~(\ref{qsedef}) of the
QSE is the relation~(\ref{Ddef}), derived in the
quenched case. In Ref.~\cite{DiGia1}, the results obtained in the
unquenched case are quite different: the exponential form of $D$
remains the same, but now $D_{1}$ is small enough to be neglected. This
approximation leads us, after the same calculations as before, to
\begin{equation}\label{qsedef2}
\Delta m^{2}=-\frac{3a}{\pi}\eta(\epsilon).
\end{equation}
The two formulas~(\ref{qsedef}) and (\ref{qsedef2}) only differ by a
constant factor: $4$ in the quenched case, $3$ in the unquenched one.
Since this factor does not seem to be presently known with a great
accuracy, we will finally use the following expression for the quark
self energy
\begin{equation}
\Delta m^{2}=-\frac{fa}{\pi}\eta(m/\delta),
\end{equation}
with $f\in\left[3,4\right]$ and $\delta\in\left[1.0,1.3\right]$~GeV.

\subsection{Insertion of QSE in effective meson models}
\label{sec:qse_in}

\par In the previous section, we showed how the QSE contribution acts as
a
shift of the squared mass of the quarks. We have now to insert this new
term in the models we described in Secs.~\ref{sec:rs_rft} and
\ref{sec:str_pert}. If we make the
substitution $m^{2}_{i}\rightarrow m^{2}_{i}+\Delta m^{2}_{i}$ in the
Hamiltonian (\ref{QCD_eq1}), we find
\begin{equation}
H\rightarrow H+\Delta H_{\text{QSE}},
\end{equation}
where
\begin{equation}\label{dHqse}
\Delta H_{\text{QSE}} = \sum^{2}_{i=1}  \frac{\Delta
m^{2}_{i}}{2\mu_{i}}
= -\frac{fa}{\pi}\sum^{2}_{i=1}\frac{\eta(m_i/\delta)}{2\mu_{i}}.
\end{equation}
Equation (\ref{dHqse}) is the total contribution of the QSE to the RS
Hamiltonian. This term has to be considered as a perturbation of the
original Hamiltonian, and thus one has not to give much sense to the
fact that for light quarks the total mass $m^{2}+\Delta m^{2}$ is
negative \cite{Sim1}. Since the QSE is a perturbation of the
Hamiltonian, it has not to be included in the elimination of the
auxiliary field $\mu$. To take into account the QSE in the RFTM, we
suggest the following procedure, inspired from
Ref.~\cite{Sim1}:
\begin{enumerate}
\item To find the eigenvalues and eigenfunctions of the RFTM
Hamiltonian.
\item To compute the mean value $\langle \mu_0 \rangle$ of the extremal
field $\mu_0$ with the eigenfunctions.
\item To add to each eigenvalue the corresponding QSE
contribution~(\ref{dHqse}) which reads, in the symmetrical case,
\begin{equation}\label{dHqses}
\Delta M_{\text{QSE}}=-\frac{fa}{\pi}\frac{\eta(m/\delta)}{
\langle \mu_0 \rangle}.
\end{equation}
\end{enumerate}

The problem is to choose the value $\mu_0$ of the extremal field. Within
the PFT, this value is given by Eq.~(\ref{muss}), and the resulting QSE
correction is given by
\begin{equation}
\label{qse_pertu}
\Delta M^{\text{PFT}}_{\text{QSE}}=
-\frac{fa}{\pi}\frac{\eta(m/\delta)}{\langle
\sqrt{\vec p\,^2+m^{2}}\rangle},
\end{equation}
as it is done in Ref.~\cite{bada02}.
On the other hand, if the equations of the RFTM are not treated in
perturbation, it seems natural to take $W_{r}\gamma_{\bot}$ for $\mu_0$
(see Sec.~\ref{sec:rs_rft}). Within this framework, the QSE contribution
is expected to be given by
\begin{equation}
\label{dHqse_rft}
\Delta M^{\text{RFT}}_{\text{QSE}}=-\frac{fa}{\pi}\frac{\eta(m/\delta)}{
\langle W_{r}\gamma_{\bot}\rangle}.
\end{equation}
Actually, we have to replace $W_{r}\gamma_{\bot}$ by
$\left\{W_r,\gamma_{\bot}\right\}/2$ in order to keep the operator
hermitian.
We can expect that both procedures will lead to different results, as in
the case of the contribution of the string as a perturbation (see
Sec.~\ref{sec:str_pert}).

\subsection{Regge trajectories}
\label{sec:Reg_t}

As mentioned in Ref.~\cite{Sim1}, the QSE correction preserves the Regge
trajectories. We can qualitatively understand this thanks to the mass
formula~(\ref{massform}), to which we add the QSE
contribution~(\ref{dHqse}). We have
\begin{equation}\label{massform2}
M^{\text{A}}(\mu_0)=4\mu_0-\frac{a\ell(\ell+1)}{16\mu^3_0}-\frac
{fa}{\pi}\frac{\eta(m/\delta)}{\mu_0}.
\end{equation}
For large angular momentum, $\mu_{0}$ becomes large. Keeping only
the dominant terms, we find the approximate mass formula
\begin{equation}
\left(M^{\text{A}}\left(\mu_{0}\right)\right)^{2}\approx 16 \mu^{2}_{0}-
8 \frac{fa}{\pi}\eta(m/\delta).
\end{equation}
It appears that the Regge trajectories are preserved, since the QSE only
causes a global shift of the squared masses and preserves the dominant
$16 \mu^2_0$
term, which grows like $\ell$. This is particularly clear when $n=0$
(see Eq.~(\ref{mumassform})).

\section{Adding the QSE}
\label{sec:qse_add}

In this section, we compute numerically the contributions of the QSE for
both the RFTM and the PFTM. We will focus on the massless
case, for which we expect the largest effect since $\eta(0)=1$ and the
extremal field $\mu_0$ is minimal. In this special case, we are
able to perform a universal analysis of our results, since the meson
mass are then just scaled by the factor $\sqrt{a}$. We will use in this
section
$f=3.0$, which is the value computed in the unquenched case.

\begin{figure}[ht]
\begin{center}
\includegraphics*[width=8.0cm]{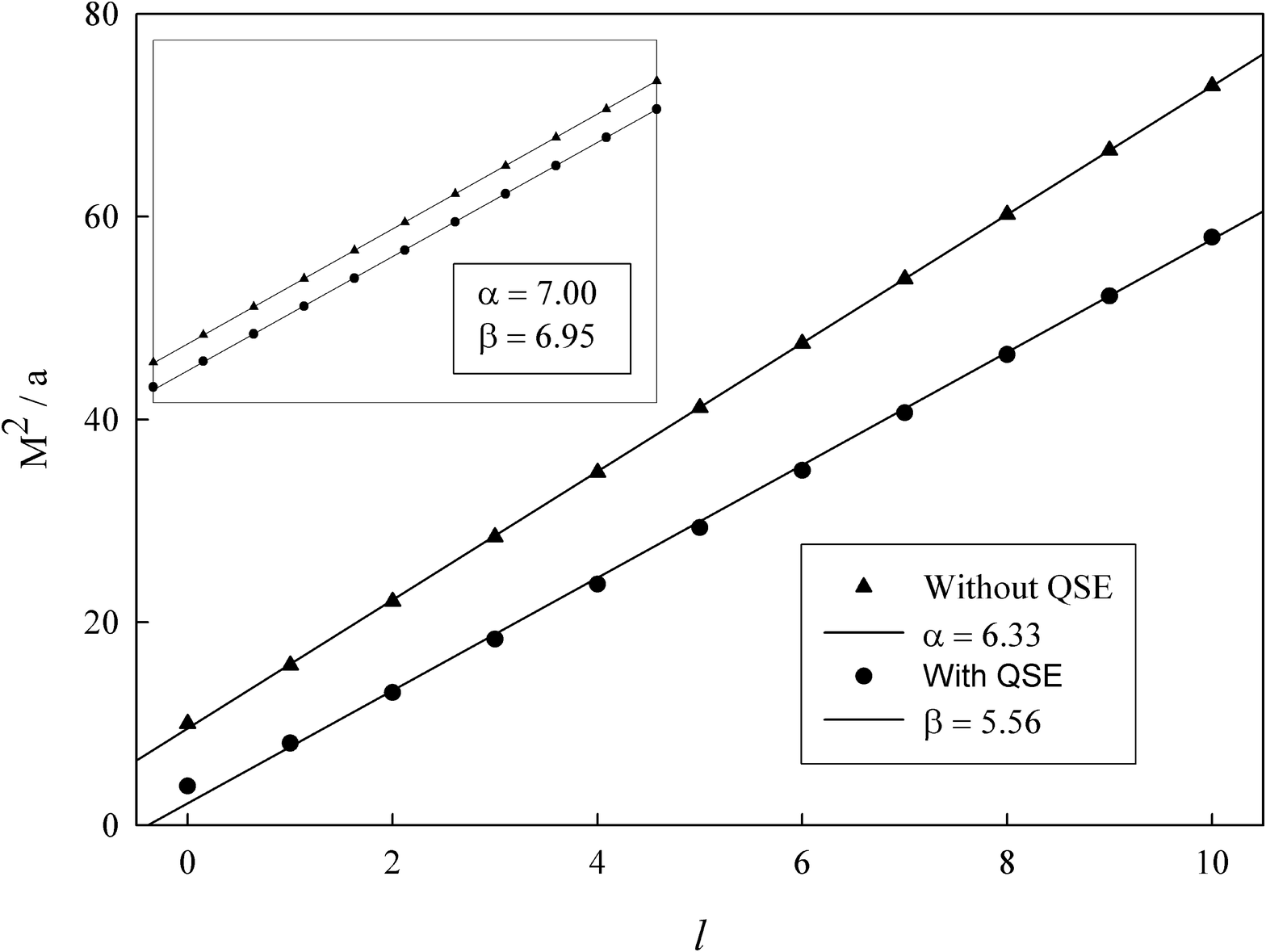}
\end{center}
\caption{Main : Regge trajectories for $n=0$ with (circle) and
without (triangle) QSE, computed with the RFTM. Small box:  Same
Regge trajectories computed with the PFTM. Lines are used to guide the
eyes.}
\label{fig:Qse1}
\end{figure}

We noticed in Sec.~\ref{sec:Reg_t} that adding the QSE contribution did
not destroy the Regge trajectories and the Regge slope. This result was
obtained using a mass formula, itself an approximation of the PFTM.
Since
we are able to numerically solve the RFTM without making approximations
\cite{buis041}, we can directly study the influence of the QSE
correction on the Regge trajectories. The main graph of
Fig.~\ref{fig:Qse1} immediately shows the negative shift of the squared
masses when
the QSE is added, as expected. If the linearity of the Regge
trajectories is
well preserved with the RFTM, the Regge slope is
not. In this figure, $\beta$ is the Regge slope with QSE and $\alpha$ is
the corresponding one without QSE. Both are rather different, and we
obtain $\beta / \alpha =0.88$. This diminution of the slope is caused by
the QSE term (\ref{dHqse_rft}). The small box shows that when one is
working with the PFTM and the QSE term (\ref{qse_pertu}), as it is the
case in \cite{bada02}, the Regge slope is not affected, or a very little
bit. We find indeed $\alpha =7.00$ and $\beta=6.95$.

\begin{figure}[h]
\begin{center}
\includegraphics*[width=8.0cm]{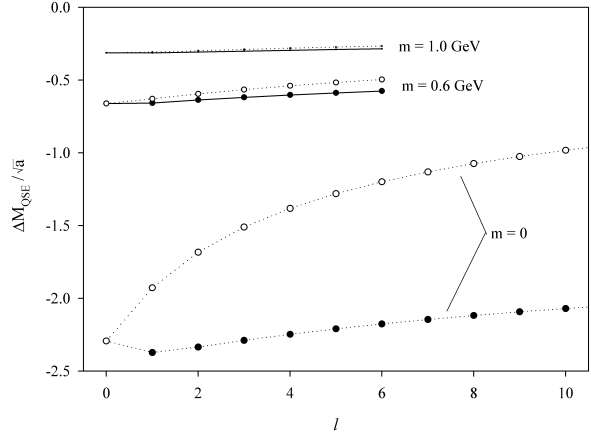}
\end{center}
\caption{QSE contributions, for $n=0$, versus $\ell$. The QSE
contributions in the RFTM (filled circle) and with the PFTM (open
circle)
are computed for different quark masses. Note that when $m\neq0$, the
scaled results are not
independent of $a$, which is taken here equal to $0.19$ GeV$^{2}$.}
\label{fig:Qse2}
\end{figure}

\begin{figure}[t]
\begin{center}
\includegraphics*[width=8.0cm]{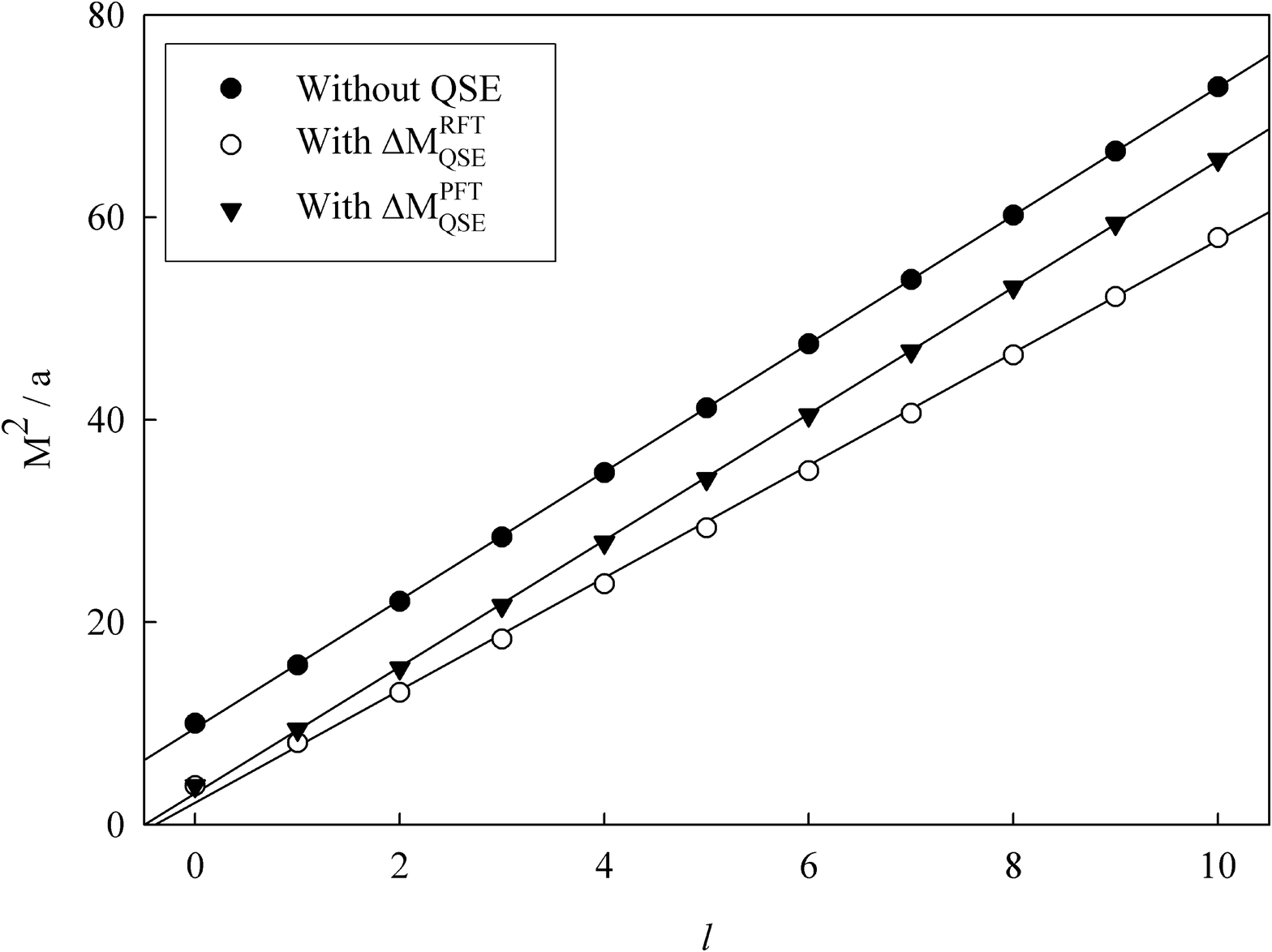}
\end{center}
\caption{Regge trajectories, with $n=0$, for the genuine RFTM (filled
circle), the RFTM with a theoretically expected QSE term
$\Delta M^{\text{RFT}}_{\text{QSE}}$ (open circle) and with a QSE term
inspired
by the PFTM $\Delta M^{\text{PFT}}_{\text{QSE}}$ (triangle).}
\label{fig:Qse3}
\end{figure}

We have now to understand why the QSE affects the Regge slope in
the RFTM and not in the PFTM.
Fig.~\ref{fig:Qse2} illustrates the differences between the QSE
contributions (\ref{qse_pertu}) and (\ref{dHqse_rft}). When
$\ell=0$ both contributions are equal since the PFTM and the RFTM reduce
to the same spinless Salpeter equation. In the massless case,
the two contributions considerably differ for $\ell \neq0$. The QSE
coming from the RFTM is
always the smaller one, and this causes the Regge slope to be smaller
when one solves the RFTM. As expected, the difference between the
two contributions decreases when the quark mass increases.
With a large quark mass, both RFTM and PFTM have a common non
relativistic limit (when the approximation $v_{\bot}\ll 1$ is the most
justified). The conclusion to draw from Fig.~\ref{fig:Qse2} is that
$\langle W_{r}\gamma_{\bot}\, \rangle$ is not even approximately equal
to $\langle \sqrt{\vec p\,^2+m^{2}}\rangle$ for light quarks.
Moreover, the second expression leads to a QSE term which preserves the
Regge slope, while the first one does not. We show in
Fig.~\ref{fig:Qse3} the effect of adding to the RFTM solutions a
``theoretically justified" QSE term (\ref{dHqse_rft}) and a ``PFTM-like"
one (\ref{qse_pertu}). As expected, the contribution (\ref{qse_pertu})
causes no diminution of the Regge slope.

\section{Experimental data}
\label{sec:exp_dat}

Since the RFTM includes neither the spin ($S$) nor the isospin ($I$) of
the mesons, the experimental data we will try to reproduce here are the
spin and isospin averaged masses, denoted $M_{\text{av}}$. These are
given by \cite{Brau98}
\begin{equation}
M_{\text{av}}=\frac{\sum_{I,J}(2I+1)(2J+1)
M_{I,J}}{\sum_{I,J}(2I+1)(2J+1)},
\end{equation}
with $\vec J=\vec L+\vec S$ and $M_{I,J}$ are different masses of the
states with the same
orbital angular momentum $\ell$. The first three
columns of Table~\ref{tab:results} show the experimental data
concerning $n\bar n$ ($n$ means $u$ or $d$) states used to compute the
spin-isospin averaged masses.
These data are taken from Ref.~\cite{pdg}.

Realistic masses cannot be computed with the genuine flux tube
Hamiltonian~(\ref{tf_equa2}). It can indeed give the right Regge slope
with $a\lesssim 0.2$ GeV, but the absolute values of the masses are
always too high. An attractive Coulomb potential, simulating the
one-gluon exchange process, must be added
\begin{equation}
V(r)=-\frac{4}{3}\frac{\alpha_{S}}{r},
\end{equation}
where $\alpha_S$ is the strong coupling constant. Theoretical arguments
as well as lattice calculations agree with a value $\alpha_S\leq 0.4$
\cite{Sim2}. However, even such a Coulomb term does not shift the masses
enough to reproduce the data (see for example Ref.~\cite{sema95}),
except if unrealistic too high values are chosen for $\alpha_S$. This
means that other contributions coming from neglected physical mechanisms
are still needed. The QSE seems to be an interesting one, because its
contribution to the mass is negative, and rather large. This should
allow to reproduce the experimental data with an acceptable value for
the Coulomb term.

\begin{table}[h]
\caption{Two sets of possible physical parameters.}
\label{tab:params}
\begin{ruledtabular}
\begin{tabular}{lcc}
  & Type $1$ & Type $2$ \\
\hline
$a$ (GeV$^{2}$) & $0.245$ & $0.195$\\
$m_{n}$ (GeV) & $0.0$ & $0.073$\\
$\alpha_{S}$ & $0.4$ & $0.4$\\
$f$ & $3.0$ & $3.0$\\
$\delta$ (GeV)& $1.0$ & $1.0$\\
QSE term & $\Delta M^{\text{RFT}}_{\text{QSE}}$ &
$\Delta M^{\text{PFT}}_{\text{QSE}}$\\
\end{tabular}
\end{ruledtabular}
\end{table}

In Table~\ref{tab:params}, we give the two sets of parameters we use
to compute the masses of some $n\bar n$ states. Both have the same
value of $f$, $\delta$ and $\alpha_S$ but differ for the other
parameters. In the type $1$, we make the usual choice $m_n=0$ and take
formula~(\ref{dHqse_rft}) as QSE term. This is the contribution that one
can theoretically expect. As we have seen that it causes a diminution of
the Regge slope, we have to take $a=0.245$ GeV$^{2}$ in order to obtain
a final slope in agreement with the experiment. Choosing $a=0.2$
GeV$^2$ is no longer possible as it is the case with the PFTM
\cite{bada02}. This is the unconventional aspect of type~1 set of
parameters. On the other side, with the type~2 set, it is possible to
keep for $a$ the standard value, about $0.2$ GeV$^2$. The price to pay
is to take formula~(\ref{qse_pertu}) as QSE term, an only ``empirically"
justified choice, and to give a small mass to the quark $n$. The
comparison between the experimental averaged masses and our results is
given in Table~\ref{tab:results}. We see that both types lead to masses
close to the spin-isospin averaged ones, our results being located
inside the error bar in almost every case.

\begin{table}[ht]
\caption{Comparison between the spin averaged masses $M_{\text{av}}$ of
the $n\bar n$ states and the results of the RFTM plus a QSE term.
$M_1$ and $M_2$ are the masses computed with the set of parameters
1 and 2 respectively. Masses are given in GeV. The first three
columns show the different states used to compute the spin averaged
masses.}
\label{tab:results}
\begin{ruledtabular}
\begin{tabular}{lccccc}
State & $I$ & $(n+1)^{2S+1}L_{J}$ & $M_{\text{av}}$ & $M_{1}$ & $M_{2}$
\\
\hline
$\omega$ & $0$ & $1^{3}S_{1}$ & $0.773\pm0.011$ & $0.788$ & $0.772$\\
$\rho$ & $1$ &  $1^{3}S_{1}$ & & &\\
\hline
$h_{1}(1170)$ & $0$ & $1^{1}P_{1}$ & $1.265\pm0.011$ & $1.269$ & $1.282$
\\
$b_{1}(1235)$ & $1$ & $1^{1}P_{1}$ & & &\\
$f_{1}(1285)$ & $0$ & $1^{3}P_{1}$ & & &\\
$a_{1}(1260)$ & $1$ & $1^{3}P_{1}$ & & &\\
$f_{2}(1270)$ & $0$ & $1^{3}P_{2}$ & & &\\
$a_{2}(1320)$ & $1$ & $1^{3}P_{2}$ & & &\\
\hline
$\omega(1650)$ & $0$ & $1^{3}D_{1}$ & $1.676\pm0.012$ & $1.673$ &
$1.678$\\
$\rho(1700)$ & $1$ & $1^{3}D_{1}$ & & &\\
$\omega_{3}(1670)$ & $0$ & $1^{3}D_{3}$ & & &\\
$\rho_{3}(1690)$ & $1$ & $1^{3}D_{3}$ & & &\\
\hline
$f_{4}(2050)$ & $0$ & $1^{3}F_{4}$ & $2.015\pm0.012$ & $2.016$ & $2.006$
\\
$a_{4}(2040)$ & $1$ & $1^{3}F_{4}$ & & &\\
\end{tabular}
\end{ruledtabular}
\end{table}

\section{Concluding remarks}
\label{sec:conclu}

The purpose of this work was to study the contribution of the quark self
energy to the meson masses in the framework of the relativistic flux
tube model. The equivalence between this phenomenological model and the
more QCD based rotating string Hamiltonian is used as a guide to perform
the calculations.

The equations defining the relativistic flux tube model being rather
complicated to solve, it seems interesting to treat the flux tube
contribution as a perturbation. Two procedures have been studied: To
eliminate firstly the auxiliary field from the rotating string
Hamiltonian, or to make firstly the approximation of small transverse
velocities in the rotating string Hamiltonian. We arrived in
Sec.~\ref{sec:str_pert} at two non equivalent terms for the flux tube
correction: the first one, obtained in the auxiliary field formalism,
was already known \cite{bada02}, and the second, obtained directly from
the relativistic flux tube equations, is a new one. The results of both
contributions are quite similar, but we showed in Sec.~\ref{sec:compar}
that the first approach is globally the best one.

Starting from the rotating string Hamiltonian and considering the
auxiliary field associated with the quark mass $\mu$ as a simple number,
an approximate but analytical mass formula is established. It enables
to understand at least qualitatively why the quark self energy preserves
the linearity of the Regge trajectories and decreases the squared masses
by a constant quantity (see Secs.~\ref{sec:massform} and
\ref{sec:Reg_t}).

The addition of the quark self energy to the relativistic flux tube
model (Sec.~\ref{sec:qse_add}) preserves the linearity of the Regge
trajectories. But, for massless quarks, the Regge slope is smaller by a
factor $0.88$ with the quark self energy than without it (this value
tends toward unity when the quark masses increase). This effect
does not exist when one works within the framework of a perturbation
theory, in which the relativistic flux tube Hamiltonian reduces to a
spinless Salpeter Hamiltonian. It is due to the fact that different
extremal values of the field $\mu$ are found, $\sqrt{\vec p\,^2+m^2}$
or $W_{r}\gamma_{\bot}$, considering the perturbation scheme or not.

In the framework of the relativistic flux tube model, it is possible to
reproduce the experimental data with the theoretically expected quark
self energy term ($\mu=W_{r}\gamma_{\bot}$) and an realistic Coulomb
term, if larger value than usual, $0.245$~GeV$^2$, is chosen for the
string tension $a$. It is possible to keep the well known value
$a=0.2$~GeV$^2$ by using a quark self energy term coming from a
perturbation approach ($\mu=\sqrt{\vec p\,^2+m^2}$). This leads us to
two possible conclusions. Within the relativistic flux tube model:
\begin{itemize}
\item It is necessary to use a quark self energy term in which the
extremal field $\mu$ is given by $\sqrt{\vec p\,^2+m^2}$ in order to
keep a value of the string tension around 0.2~GeV$^2$.
\item It is necessary to use a quark self energy term in which the
extremal field $\mu$ is given by $W_{r}\gamma_{\bot}$, the natural value
associated with the relativistic flux tube model; But another physical
mechanism has to be taken into account in order to keep a value of the
string tension around 0.2~GeV$^2$.
\end{itemize}
It seems hard to justify the first proposal by any theoretical argument.
The mentioned new mechanism could be due to retardation effects or
deviations from the straight line for the flux tube. Such a work is in
progress.

\acknowledgments

C.~S. (FNRS Research Associate) and F.~B. (FNRS Research
Fellow) thank the FNRS for financial support.


\begin{thebibliography}{99}

\bibitem{laco89} D. LaCourse and M. G. Olsson, Phys. Rev. D {\bf 39},
2751 (1989).
\bibitem{tf_2} M. G. Olsson and S. Veseli,\ Phys. Rev. D\ \textbf{51},\
3578\ (1995).
\bibitem{dubi94} A. Yu. Dubin, A. B. Kaidalov, and Yu. A. Simonov, Phys.
Lett. B \textbf{323}, 41 (1994) [hep-ph/9311344].
\bibitem{morg99} V. L. Morgunov, A. V. Nefediev, and Yu. A. Simonov,
Phys. Lett. B \textbf{459}, 653 (1999) [hep-ph/9906318].
\bibitem{sema04} C. Semay, B. Silvestre-Brac, and I. M. Narodetskii,
Phys. Rev. D {\bf 69}, 014003 (2004) [hep-ph/0309256].
\bibitem{buis042} F. Buisseret and C. Semay, Phys. Rev. D \textbf{70},
077501 (2004) [hep-ph/0406216].
\bibitem{sema95} C. Semay and B. Silvestre-Brac, Phys. Rev. D
{\bf 52}, 6553 (1995).
\bibitem{Sim1} Yu. A. Simonov, Phys. Lett. B \textbf{515}, 137 (2001).
\bibitem{guba94} E. L. Gubankova and A. Yu. Dubin, Phys. Lett.
B \textbf{334}, 180 (1994) [hep-ph/9408278].
\bibitem{buis041} F. Buisseret and C. Semay [hep-ph/0409033].
\bibitem{bada02} A. M. Badalian and B. L. G. Bakker, Phys. Rev. D
\textbf{66}, 034025 (2002) [hep-ph/0202246].
\bibitem{luch95} W. Lucha and F. F. Sch\"oberl, Phys. Rev. A
{\bf 51}, 4419 (1995).
\bibitem{bose76} S. K. Bose, A. Jabs, and H. J. W. M\"uller-Kirsten,
Phys. Rev. D {\bf 13}, 1489 (1976).
\bibitem{feyn1} R. P. Feynman, Phys. Rev. \textbf{56}, 340 (1939).
\bibitem{DiGia1} A. Di Giacomo and Yu. A. Simonov, Phys. Lett. B \textbf
{595}, 368 (2004).
\bibitem{DiGia2} A. Di Giacomo and H. Panagopoulos, Phys. Rev. B \textbf
{285}, 133 (1992).
\bibitem{Brau98} F. Brau and C. Semay, Phys. Rev. D \textbf{58}, 034015
(1998).
\bibitem{pdg}Particle Data Group, S. Eidelman \textit{et al.}, Phys.
Lett. B \textbf{592}, 1 (2004).
\bibitem{Sim2} Yu. A. Simonov, JETP Lett. \textbf{57}, 525 (1993).

\end{thebibliography}
\end{document}